\documentclass[letterpaper,english,aps,prl,reprint,superscriptaddress]{revtex4-2}
\usepackage[T1]{fontenc}
\usepackage[utf8]{inputenc}
\usepackage{color}
\usepackage{amsmath}
\usepackage{amssymb}
\usepackage{mathrsfs}
\usepackage{mathtools}
\usepackage{bm}
\usepackage{graphicx}
\usepackage{extarrows}
\usepackage{amsthm}

\usepackage[most]{tcolorbox}
\newtcbox{\othermathbox}[1][]{nobeforeafter, math upper, tcbox raise base, 
          enhanced, rounded corners, colback=black!5, colframe=black,
          left=0.7em, top=0.4em, right=0.7em, bottom=0.5em}
\usepackage{empheq}

\definecolor{MyYellow}{RGB}{248,199,82}

\usepackage{titlesec}
\usepackage{enumerate}

\let\OLDthebibliography\thebibliography
\renewcommand\thebibliography[1]{
  \OLDthebibliography{#1}
  \setlength{\parskip}{0pt}
  \setlength{\itemsep}{3pt plus 0.3ex}
}

\usepackage[colorlinks,allcolors=blue]{hyperref}

\definecolor{LightBrown}{RGB}{255,188,0}
\definecolor{MiddleBrown}{RGB}{199,146,0}
\definecolor{DarkBrown}{RGB}{143,104,0}
\definecolor{DarkerBrown}{RGB}{87,62,0}
\definecolor{Purple}{RGB}{255,0,188}

\newcommand{\be}{\begin{equation}}
\newcommand{\ee}{\end{equation}}
\renewcommand{\bm}{\begin{pmatrix}}
\renewcommand{\em}{\end{pmatrix}}

\renewcommand{\zeta}{w}

\makeatletter

\pdfpageheight\paperheight
\pdfpagewidth\paperwidth

\usepackage{hyperref}
\hypersetup{
    colorlinks = true,
	allcolors = blue	
}
\usepackage{times}

\makeatother

\usepackage{babel}
\begin{document}
\title{Emergent Self-Similar Quantum Revivals in Spiral Drives}
\author{Xin-Chi Zhou}
\thanks{These authors contributed equally to this work.}
\affiliation{Max Planck Institute for the Physics of Complex Systems, N\"othnitzer Stra\ss e 38, 01187 Dresden, Germany}
\author{Liang-Hong Mo}
\thanks{These authors contributed equally to this work.}
\affiliation{Department of Physics, Princeton University, Princeton, New Jersey 08544, USA}
\author{Hongzheng Zhao}
\thanks{hzhao@pku.edu.cn}
\affiliation{State Key Laboratory of Artificial Microstructure and Mesoscopic Physics, School of Physics, Peking University, Beijing 100871, China}
\author{Bastien Lapierre}
\thanks{bastien.lapierre@phys.ens.fr}
\affiliation{Philippe Meyer Institute, Physics Department, \'Ecole Normale Sup\'erieure (ENS), Université PSL, 24 rue Lhomond, F-75231 Paris, France}
\begin{abstract}
We uncover a distinct form of nonequilibrium temporal order: self-similar quantum revivals in a many-body system driven  by quasiperiodic spiral kicks, where the system recurrently returns close to its initial state at a hierarchically nested sequence of times. We demonstrate that both the fidelity and entanglement entropy exhibit this self-similar temporal structure. It originates from an emergent dynamical attractor, which we identify, such
that all momentum modes eventually fall into the same closed
orbits at self-similar times.
We analytically justify this behavior and show that, for special momentum modes, this attractor arises as a consequence of a generalized spin echo process, and more generally we prove its existence  using quasiperiodic SU(2) cocycles. Interestingly, the dynamics between consecutive revivals supports either volume- or area-law entanglement scaling, tunable via the driving parameters.  In the presence of integrability-breaking perturbations, the system eventually heats up, but a long-lived prethermal regime with algebraically tunable lifetime occurs before heating sets in. Our results establish self-similar quantum revivals as a new paradigm for nonequilibrium quantum matter and provide a realistic route for its observation in current quantum simulators.
\end{abstract}
\maketitle

\textit{Introduction}.--
Spatial self-similarity plays an important role in condensed matter physics, most prominently in quasicrystals, where deterministic yet quasiperiodic order recurs across multiple length scales~\cite{PhysRevLett.53.2477, PhysRevLett.53.1951}. 
This leads to fractal structures in the spectrum~\cite{RevModPhys.93.045001}, together with exotic critical eigenstates exhibiting fractal spatial fluctuations~\cite{Kohmoto1987CriticalWaveFunctions}.
This hierarchical organization has greatly broadened the conventional paradigm of equilibrium phases beyond crystalline order~\cite{Kraus2016Quasiperiodicity,Wang2021ManyBodyCritical,
Goncalves2024QuasiFractalMoire}. 

A natural question is whether self-similarity can also emerge in the time domain as a fundamental organizing principle of far-from-equilibrium quantum matter. 
 Time-dependent driving provides one promising way to address this question~\cite{Bukov2015UniversalHighFrequency,Khemani2016PhaseStructure, Else2016FloquetTimeCrystals,
Moessner2017FloquetMatter,
Harper2020FloquetTopology}. For instance, using periodic~\cite{Autti2018TimeQuasicrystal,Flicker2018TimeQuasilattices,giergiel2019Discrete,
pizzi2019Periodn,
Chinzei2020FloquetDynamicalSymmetry} or aperiodic
driving~\cite{PhysRevLett.120.070602,zhao2019Floquet,
else2020LongLived,moon2025experimental,
he2025Experimental,Fang2025QuasiperiodicPhaseTransitions,
Zhu2025RydbergTimeQuasicrystal,Mo2025ComplexHeating,schmid2025SelfSimilar,Eckstein2026SelfDualCriticality}, local observables of many-body systems can exhibit discrete time quasi-crystalline order, which spontaneously breaks the discrete time-translation symmetry of the external drive. While generic interacting systems can absorb energy from the drive and eventually heat up to a featureless state~\cite{DAlessio2014LongTime,
Lazarides2014Equilibrium,Ponte2015ErgodicMBL}, high-frequency driving~\cite{Abanin2015ExponentiallySlowHeating,Mori2016EnergyAbsorption,
Abanin2017EffectiveHamiltonians,
Abanin2017RigorousPrethermalization,
Else2017PrethermalPhases} or strong disorder-induced localization~\cite{Ponte2015ManyBodyLocalization,
Lazarides2015FateMBL,
Abanin2016DrivenMBLTheory,
Khemani2016PhaseStructure,
Else2016FloquetTimeCrystals,
vonKeyserlingk2016AbsoluteStability,Zhang2017DiscreteTimeCrystal,Choi2017TimeCrystallineOrder,Yao2017DiscreteTimeCrystals,long2022Manybody,Zaletel2023TimeCrystalsColloquium} can be exploited to suppress heating, thereby stabilizing nonequilibrium phases of matter~\cite{PhysRevX.7.031034,Peng2018MajoranaMultiplexing,crowley2019Topological,boyers2020Exploring,crowley2020HalfInteger, PhysRevResearch.2.033461,Long2021TopologicalPumps,PilatowskyCameo2025MorphicDrives,Wu2026GeometricQuantumDrives}.

In this Letter, we uncover a qualitatively distinct form of nonequilibrium self-similar temporal order: a driven many-body system exhibiting self-similar quantum revivals, in which the wavefunction recurrently returns close to its initial state at a hierarchically nested sequence of times. This can be achieved by considering a family of quasiperiodic spiral drives, where a clean Ising chain experiences irrational transverse `kicks' at discrete time steps.
Crucially, the characteristic revival does not rely on high-frequency driving or on localization; rather, it relies on an emergent dynamical attractor, which we identify, that appears asymptotically at long times, preventing the system from evolving ergodically.

We begin with an integrable setting, which allows for efficient long-time and large-scale numerical simulation, and explicitly show that both the state fidelity and entanglement entropy exhibit the characteristic self-similar temporal order. In momentum space, the system's evolution is governed by SU(2) transfer matrices and the dynamical attractor appears, such that all momentum modes eventually fall into the same closed orbits at self-similar times. We justify this behavior analytically and for pedagogical reasons, we start from special modes for which the attractor arises as a consequence of a generalized spin echo process, where deviations away from the initial state are echoed out at self-similar times; then we prove its existence in the general case
assuming reducibility properties of quasiperiodic cocycles over the SU(2) group~\cite{Krikorian2001GlobalDO, avila2007reducibilitynonuniformhyperbolicityquasiperiodic}.

Interestingly, between two consecutive revivals, the spiral drive can generate either volume- or area-law entanglement entropy scaling tunable via driving parameters. The entanglement growth is also sharply different in these two regimes: the former exhibits standard linear-in-time entanglement growth, while the latter features logarithmically slow growth. 

In the presence of generic integrability-breaking perturbations, the system eventually heats up towards an infinite temperature state. However, crucially, a long-lived prethermal regime appears with an algebraically tunable lifetime before the eventual heat death. Our work greatly enriches the zoo of nonequilibrium phases of matter with self-similar temporal order and provides a realistic route for its realization in state-of-the-art quantum simulators.

\begin{figure*}[htp!]
\centering
\includegraphics{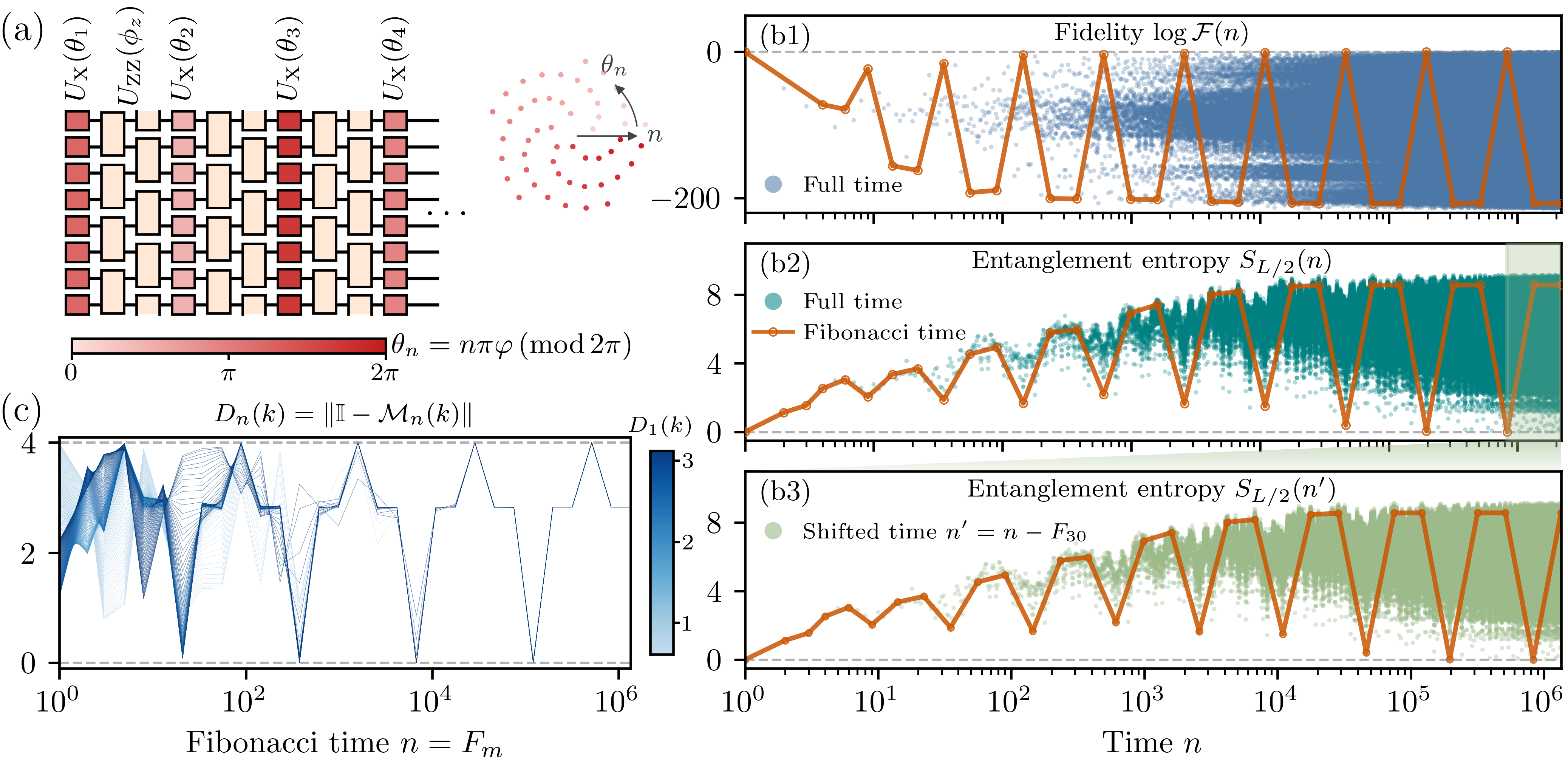}
\caption{\label{Fig1} 
(a) Circuit representation of the spiral drive introduced in Eq.~\eqref{eq:mosaic_setup}. Different colors for the X gates indicate different irrational single-qubit rotations, following $\theta_n = n \pi \varphi$ $(\mathrm{mod}$ $2\pi$). These successive rotations form a characteristic Fermat's spiral. (b) Time evolution of the fidelity $\mathcal{F}(n)$ and half-system entanglement entropy $S_{L/2}(n)$ as a function of the stroboscopic step $n$, for $L=1000$, and $\phi_z=\pi/3$. The fidelity reveals near revivals at special times, highlighted in orange, corresponding to $n=F_{3m}$. 
Similar features can be observed in $S_{L/2}(n)$, which grows logarithmically with $n$ before eventually saturating to low-entanglement states. A magnified view of $S_{L/2}(n')$ starting at $n'=n-F_{30}$ shows the manifest self-similarity of the time evolution. (c) The time evolution of any given coherent state~\eqref{eq:coherentstate} is encoded in quasiperiodic products of SU(2) matrices, denoted as $\mathcal{M}_n(k)$ for each $k$ mode. We show the Frobenius norm to the identity, $\|\mathbb{I}-\mathcal{M}_n(k)\|$, evaluated at Fibonacci times $n=F_m$ for each $k$. Each individual trajectory, color-coded by its initial value $\|\mathbb{I}-\mathcal{M}_1(k)\|$, converges to a unique asymptotic trajectory, such that the product converges to the identity (up to an overall sign) at steps $n=F_{3m}$. 
}

\end{figure*}

\textit{Spiral drive.}--We introduce a quasiperiodically driven unitary circuit, termed the \textit{spiral} drive [Fig.~\ref{Fig1}(a)], whose evolution after $n$ layers is given by
\begin{equation}
\label{eq:mosaic_setup}
U(n)=\prod_{m=1}^{n}\Big[U_{\rm ZZ}(\phi_z) U_X(\theta_{m})\Big],
\end{equation}
where the Ising gate $U_{\rm ZZ}$ and the global rotation $U_X$ read
\begin{equation*}
    U_{\rm ZZ}(\phi_z)=e^{-i \phi_z \sum_{j=1}^L Z_{j}Z_{j+1}}, \quad U_X(\theta_n)=e^{-i\theta_n \sum_{j=1}^L X_j}.
\end{equation*}
Here, $Z_j$ and $X_j$ are Pauli operators acting on the $j$-th qubit, and periodic boundary conditions are imposed. 
The key feature of the spiral drive is the irrational transverse kick, whose angle $\theta_n = n\pi\varphi$ is set by the golden ratio $\varphi=\lim_{m\rightarrow\infty}F_{m-1}/F_m=(\sqrt{5}-1)/2$, with $F_m$ the $m$-th Fibonacci number. Because $\varphi$ is irrational, the sequence $\theta_n\bmod 2\pi$ never repeats and winds around the unit circle, tracing a spiral pattern in time that gives the drive its name, as shown in Fig.~\ref{Fig1}(a). By the equidistribution theorem, the sequence $\{\theta_n\bmod 2\pi\}$ covers the circle densely and uniformly. 
These properties make the drive \textit{quasiperiodic}, which is deterministic and temporally ordered, yet crucially non-periodic.
Contrary to previously studied Fibonacci drives~\cite{PhysRevLett.57.770, PhysRevLett.120.070602,Wen2021DrivenCFT,PilatowskyCameo2023CompleteErgodicity,lapierre2025entanglementtransitionsstructuredrandom}, in which two noncommuting unitary operators are arranged according to a Fibonacci word, the spiral drive is built on an irrational sequence of transverse kicks. 
The findings presented below are not limited to the golden ratio, and related results on self-similar revivals persist for generic irrational numbers, as detailed in the supplementary material (SM)~\cite{supplement}.

Both $U_{\rm ZZ}$ and $U_X$ are Gaussian-preserving
gates and, under the Jordan-Wigner transformation, map to noninteracting fermionic
operators. The many-body dynamics therefore remains within the manifold of fermionic
coherent states
\begin{equation}
\label{eq:coherentstate}
    \lvert \psi(\mathcal{N},f) \rangle
    = \mathcal{N}  e^{\sum_{k\in \Lambda_{+}} f(k)  c_{-k}^{\dagger} c_k^{\dagger}}
    |0\rangle,
\end{equation}
where $c_k^{\dagger} (c_k)$ creates (annihilates) a fermion with momentum $k$,
$\Lambda_+=(2\pi/L)\{1/2,\ldots,L/2-1/2\}$ is the set of positive momenta compatible
with antiperiodic boundary conditions, $f(k)$ is the amplitude of the pair $(-k,k)$,
$\mathcal{N}$ is the normalization, and $|0\rangle = \bigotimes_j \lvert+\rangle_j$
with $\lvert+\rangle_j$ the $+1$ eigenstate of $X_j$. Each Gaussian gate preserves
the coherent-state form~\cite{dreyer2021Quantum,granet2023VolumeLaw},
$U_{\mathcal{O}}(t)\lvert\psi(\mathcal{N},f)\rangle
 = \lvert\psi(\widetilde{\mathcal{N}},\tilde f)\rangle$
with $\mathcal{O}\in\{\mathrm{ZZ},X\}$. In each momentum sector,  the amplitude $f(k)$ is updated by the M\"obius transformation
\begin{equation}\label{Eq:mobius}
    \tilde f(k)  =  \frac{\alpha_k  f(k) + \beta_k}{-\beta_k^{*}  f(k) + \alpha_k^{*}},
\end{equation}
which can be written as $(\tilde f(k),1)^\top\propto G_{\mathcal{O}}(k,t) (f(k),1)^\top$,
where $G_{\mathcal{O}}(k,t)\in\mathrm{SU}(2)$ has its explicit form for $\mathcal{O}\in\{\mathrm{ZZ},X\}$ given in the SM~\cite{supplement}.
In each momentum sector, the time evolution is thus governed by the product of a series of SU(2) matrices, which we denote by $\mathcal{M}_n(k)$ after $n$ time steps.
The time evolution of various observables, including entanglement entropy (EE) and state fidelity, can be obtained from the correlation matrix built from $\mathcal{M}_n(k)$ (see SM~\cite{supplement}).

\textit{Emergent self-similar revivals.}--
The spiral drive defined by Eq.~\eqref{eq:mosaic_setup} generates a hierarchy of quantum revivals at Fibonacci times, where the many-body state returns arbitrarily close to its initial configuration at every third Fibonacci time step $n=F_{3m}$.
To diagnose this behavior, we track  the fidelity $\mathcal{F}(n)=|\langle\psi(0)|\psi(n)\rangle|^2$ and the half-chain bipartite EE $S_{L/2}(n)=-{\rm Tr}[\rho_{L/2}(n)\log\rho_{L/2}(n)]$, with $\rho_{L/2}(n)$ being the reduced density matrix of the left-half subsystem.
We initialize the system in $|0\rangle$ without loss of generality, 
since the results reported below hold for any initial state of the form of Eq.~\eqref{eq:coherentstate}. 
Fig.~\ref{Fig1}(b1,b2) shows that the dynamics exhibits an asymptotic period-3 structure in the Fibonacci index.
In particular, at sufficiently long times, both observables return arbitrarily close to their initial value at 
$n=F_{3m}$, where $S_{L/2}\to 0$ and $\mathcal{F}\to 1$, implying that the time-evolved state itself nearly returns to the initial state at these special times. We stress that these revivals are emergent: they do not occur at early times, and are thus a consequence of the long-time structure of the spiral drive.

Moreover, the asymptotic revival pattern itself is self-similar in time. Indeed, Fig.~\ref{Fig1}(b3) resolves $S_{L/2}(n')$ starting from the revival at $n=F_{30}=832040$, with shifted time $n'=n-F_{30}$, and reproduces the dynamical structure already observed in Fig.~\ref{Fig1}(b2). 
Iterating this construction yields, in the infinite-time limit, an infinite hierarchy of quantum revivals, beyond Fibonacci times.

\textit{Mechanism of revivals}.-- 
We now prove analytically the origin of the revivals: at self-similar times, every momentum mode falls into the same closed period-6 orbit. Following the structure in Eq.~\eqref{Eq:mobius}, a single momentum mode under the spiral drive evolves through the product of matrices 
\begin{equation}
M_n \equiv P e^{2\pi i \varphi n \sigma_z}P e^{2\pi i \varphi (n-1) \sigma_z}\cdots P e^{2\pi i \varphi  \sigma_z},
\label{eq:matrixproductSU2}
\end{equation}
where $P$ is an arbitrary non-diagonal SU(2) matrix that encodes a generic Gaussian gate acting on a given momentum mode. 
For instance, $G_{\mathrm{ZZ}}(k,\phi_z)$ is one such realization of $P$, which, together with $G_X(k,\theta_n)=e^{2i\theta_n\sigma_z}$, builds the transfer matrix $\mathcal{M}_n(k)$ for the spiral drive.
Our goal is to show that $M_{F_{3m}}$ converges to $\pm \mathbb{I}$ asymptotically for any non-diagonal $P\in\mathrm{SU}\,(2)$, which directly implies the appearance of the self-similar revival.

We begin with the case where $P$ is purely off-diagonal, for which the period-6 orbit can be derived rigorously and understood through a spin-echo picture. 
We first notice that $P$ effectively acts as a $\pi$-pulse that flips the $z$-axis of the spin $P\sigma_zP^{-1}=-\sigma_z$. 
Thus, this $\pi$-pulse naturally turns the forward precession, generated via $e^{2\pi i\varphi\sigma_z}$ around the $z$-axis, into the backward precession. Consequently,
each neighboring pair of $\pi$-pulses produces a perfect echo up to a residual rotation of angle $\varphi$.
Accumulating all the residual rotations over the full product in  Eq.~\eqref{eq:matrixproductSU2} yields the closed form $M_n=e^{2\pi i \varphi \lceil n/2\rceil \sigma_z}P^{n}$; see derivations in SM~\cite{supplement}.
This readily implies that for  $n\equiv 0\pmod 3$, $M_{F_n}$ converges to the identity matrix up to an overall sign, independently of the choice of $P$. 
Importantly, this behavior remains robust even when we slightly perturb this off-diagonal case by introducing diagonal elements of strength $\epsilon$ to the matrix $P$. As shown in SM~\cite{supplement}, the leading order $\mathcal{O}(\epsilon)$ contribution to Eq.~\eqref{eq:matrixproductSU2} that induces a deviation away from the perfect identity matrix at self-similar times also gets echoed out at long times.

Away from this limit, no closed form exists, yet the period-6 orbit can still be obtained by interpreting the product as the iteration of a quasiperiodic cocycle over SU(2), which becomes tractable when the cocycle is assumed reducible~\cite{Krikorian2001GlobalDO}. As defined in detail in End Matter, quasiperiodic cocycles are pairs $(\varphi,A)$ of a rotation $\varphi \in\mathbb R\setminus\mathbb Q$ and a smooth map $A(x)$ from the circle $\mathbb R/\mathbb Z$ to SU(2), such that an iteration acts as $(x,\psi)\mapsto (x+\varphi,A(x)\psi)$.
In our case,  we choose $A(x)=Pe^{2\pi i x\sigma_z}$ such that the $n$-th iteration of this cocycle, denoted $A_n(x)$, reduces exactly to Eq.~\eqref{eq:matrixproductSU2} for $x=\varphi$.

Under the natural assumption that the cocycle is reducible~\cite{Krikorian2001GlobalDO}\footnote{We note that the set of reducible single frequency cocycles over SU(2) is dense~\cite{Krikorian2001GlobalDO}, which justifies this assumption.}, there exists $B(x)=B(x+1)$ and a constant matrix $C\in \mathrm{SU}(2)$ such that
\begin{equation}
A(x) = B(x+\varphi)CB(x)^{-1}.
\label{eq:reducibilitydef}
\end{equation}
Iterating Eq.~\eqref{eq:reducibilitydef}, one has $A_{m}(x) = B(x+m\varphi)C^{m} B(x)^{-1}$. Thus, reducibility implies a quasiperiodic analog of Floquet's theorem: $C$ plays the role of a fixed effective rotation, while $B(x)$ gives the endpoint micromotion factors.

When evaluating $A_{m}(x)$ at Fibonacci steps $m=F_n$, the phase $x$ almost returns to its initial value, $x+F_n\varphi=x+O(\varphi^{n})$ mod 1 as a consequence of Binet's identity $F_n\varphi=F_{n-1}-(-\varphi)^n$.
The two endpoint micromotions are therefore exponentially close, giving rise to $\text{Tr}(A_{F_n}(x)) = \text{Tr}(C^{F_n})+O(\varphi^{n})$. This reduces the problem from characterizing an infinite quasiperiodic product to finding the eigenvalues of $C$.  The spectrum of this effective rotation is crucially constrained by the half-cycle symmetry of $A(x)=Pe^{2\pi i x\sigma_z}$,
\begin{equation}
A\left(x+\frac{1}{2}\right)=-A(x).
\label{eq:symmetruycocle}
\end{equation}
This symmetry, together with the reducibility assumption in Eq.~\eqref{eq:reducibilitydef}, can be used to show that the eigenvalues of $C$ are $\{e^{i\beta},e^{-i\beta}\}$ with $\beta=\pi k \varphi-\frac{\pi}{2}$ for $k\in2\mathbb{Z}+1$ (see End Matter). Therefore, $\text{Tr}(C^{F_n})\rightarrow 2\cos(-\frac{\pi F_n}{2}+\pi k F_{n-1})$, from which the period-6 cycle \{0,\,0,\,2,\,0,\,0,\,-2\} follows. 
This leads us to the conclusion that $M_{F_n}$ converges to the identity matrix at steps  $n\equiv 0\pmod 3$, independently of $P$, as long as the cocycle is reducible.

The above mechanism accounts for the revivals discussed previously: for each momentum $k$, the time evolution follows a quasiperiodic product of the form Eq.~\eqref{eq:matrixproductSU2}, such that at Fibonacci steps $F_{3m}$, each $k$-dependent product converges towards the identity matrix (up to an irrelevant overall sign), as illustrated in Fig.~\ref{Fig1}(c). Consequently, any initial coherent state asymptotically revives at these special times, in agreement with the numerical results in Fig.~\ref{Fig1}(b).
We further stress that revivals occur for spiral drives constructed from \textit{any} operator that preserves coherent states, going much beyond the specific driving protocol in Eq.~\eqref{eq:mosaic_setup}; we will return to this below. 
In this sense, the revivals are an unavoidable consequence of reducible cocycles satisfying the symmetry in Eq.~\eqref{eq:symmetruycocle}.
Importantly, this robust revival mechanism is unique to spiral drives and does not exist for other types of periodic, quasiperiodic, or random drives.

\begin{figure}[t!]
\centering
\includegraphics[scale=1]{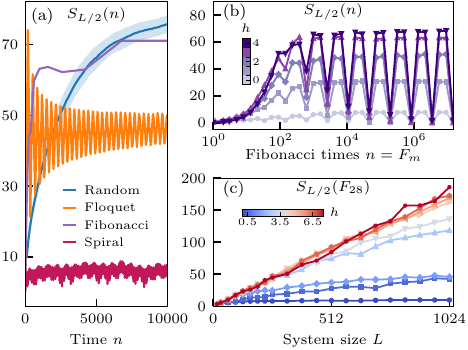}
\caption{\label{Fig2} 
(a) The bipartite entanglement entropy $S_{L/2}(n)$ for the spiral drive compared with three reference protocols: the random drive with $\theta_n$ sampled uniformly in $[0,2\pi)$ (shaded region: one standard deviation over realizations), the Floquet drive with fixed $\theta_n\equiv\theta$, and the Fibonacci drive in which $\mathrm{ZZ}$ and $\mathrm{X}$ gates alternate 
according to the Fibonacci substitution rule. 
The spiral drive 
exhibits highly reduced entanglement growth compared to any of the other protocols. The system size is $L=400$, with $h=0$ and $\phi_z=\pi/3$. 
(b) $S_{L/2}(n)$ for the spiral drive at $n=F_m$ for different transverse fields $h$ in~\eqref{eq:transversefield}. As $h$ increases, the intermediate entanglement growth goes from logarithmic to linear, while the self-similar quantum revivals persist for any value of $h$ at later times.
(c) $S_{L/2}(F_{28})$ versus total system size $L$ for $h\in\{0,0.5,1.5,2.5,\cdots,7.5\}$. As $h$ increases, the entanglement scaling is tuned from area law to volume law.} 
\end{figure}

\textit{Tunable entanglement growth}.--
We now show that the spiral drive can be tuned between two regimes, one featuring logarithmic growth toward an area-law state, and the other with linear growth toward a volume-law state, while preserving the self-similar revival.
To establish this tunability, we replace the Ising gate $U_{\rm ZZ}(\phi_z)$ in Eq.~\eqref{eq:mosaic_setup} with the transverse-field Ising (TFI) gate
\begin{equation}
\label{eq:transversefield}
U_{\rm TFI}(\phi_z,h)=e^{-i \phi_z\sum_j  \left(Z_j Z_{j+1}+h X_j\right)},
\end{equation} 
where $h$ tunes the transverse-field strength. At $h=0$, this gate reduces to the original Ising gate. As shown in Fig.~\ref{Fig1}(b2), the half-chain EE grows only logarithmically with $n$, with  numerical verification given in the SM~\cite{supplement}, before saturating at a low-entanglement plateau. 
This logarithmic growth of entanglement is anomalous: in any translation-invariant driven system, entanglement generically grows ballistically, eventually saturating to a volume-law state. The spiral drive is a striking exception to this rule, as made clear by direct comparison with periodic, random, and quasiperiodic drives in Fig.~\ref{Fig2}(a).
In Fig.~\ref{Fig2}(b), we show the time evolution of EE
at Fibonacci steps for different values of $h$ in Eq.~\eqref{eq:transversefield}. 
As $h$ increases, the early-time logarithmic growth crosses over to linear growth, and the plateau entanglement entropy between revivals approaches the maximal Gaussian Page value~\cite{Bianchi2021PageCurve}. Crucially, these near-revivals remain accessible at moderate times even for large system sizes ($L\sim10^3$), distinguishing the spiral-drive revivals from accidental finite-size recurrences in generic driven free-fermion models, which are expected to become exponentially rare with system size.

The late-time entanglement scaling exhibits an area-to-volume crossover as $h$ increases. Fig.~\ref{Fig2}(c) shows the half-chain EE at Fibonacci time $n=F_{28}$ as a function of total system size $L$. At $h=0$, the EE saturates to an $L$-independent value, consistent with the area law previously observed. A clear crossover emerges near $h\approx1.5$, above which the half-chain EE grows extensively with $L$, manifesting a volume law. 
Despite this dramatically enhanced entanglement growth, asymptotic revivals still emerge at time steps $n=F_{3m}$ for any $h$. 
This is consistent with our mathematical argument: the spiral drive built from any Gaussian-preserving gate protects the asymptotic revivals, which therefore persist even when the intermediate entanglement reaches maximal values. The coexistence of self-similar revivals with volume-law entanglement shows that the spiral drive preserves memory of the initial state, which is systematically recovered at prescribed times, even when entanglement saturates to its maximum in between.

\begin{figure}[t]
\centering
\includegraphics[width=1.0\columnwidth]{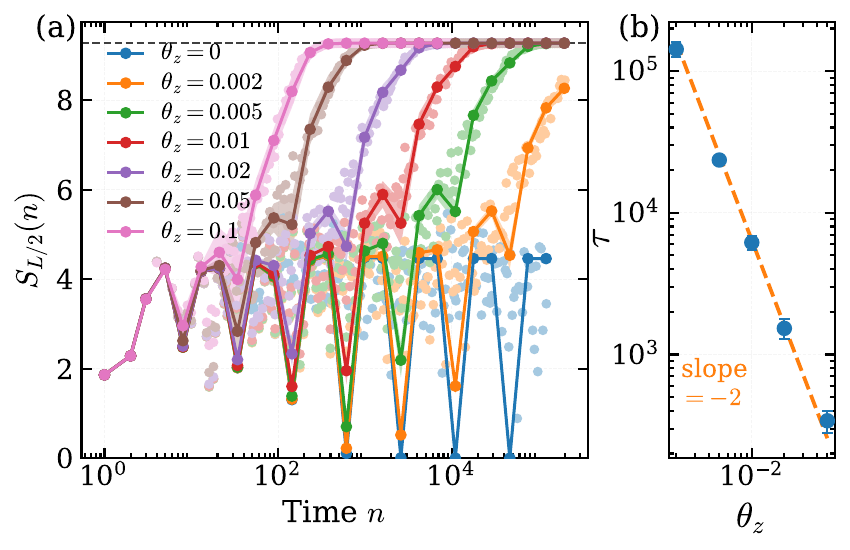}
\caption{\label{Fig3}Parametrically long-lived non-heating dynamics under random integrability-breaking. (a) The bipartite entanglement entropy $S_{L/2}(n)$ at Fibonacci times $n=F_m$ (filled circles) and at logarithmically spaced intermediate times (light dots) for different noise strengths $\theta_z$. A prethermal regime emerges, in which the non-heating dynamics persists, as evidenced by revivals toward states close to the initial state. Shaded bands indicate $\pm$ one standard deviation over $10$ random realizations, and the dashed line marks the Page value. (b) Prethermal lifetime $\tau$ versus $\theta_z$. Here $\tau$ is defined as the time at which $S_{L/2}(n)$ exceeds the threshold  $S^{c}_{L/2}=8$. The fitted scaling $\tau \sim \theta_z^{-2}$ (dashed line) demonstrates a parametrically enhanced lifetime, consistent with Fermi's golden rule. The system size is $L=20$.}
\end{figure}

\textit{Robustness against nonintegrable perturbations.}--
The self-similar 
revivals predicted above remain parametrically long-lived under generic integrability-breaking perturbations, rendering them directly observable in contemporary quantum simulators. 
To demonstrate this, we perturb the spiral drive in Eq.~\eqref{eq:mosaic_setup} with an additional longitudinal field, 
\begin{equation}
U(n)\rightarrow\prod_{m=1}^{n}\Big[U_{\rm ZZ}(\phi_z)e^{-i\delta_m\sum_j Z_j}U_X(\theta_{m})\Big],
\end{equation}
where $\delta_m=\theta_z r_m$, with $r_m$ uniformly sampled from $[-1,1]$ at each time step. 
The temporal disorder mimics the realistic noise present in actual quantum devices. The perturbation strength $\theta_z$ breaks integrability and ultimately drives the system to thermalize.
Fig.~\ref{Fig3}(a) shows the disorder-averaged EE for a range of $\theta_z$. 
Importantly, for small $\theta_z$, the system first enters a long-lived prethermal regime where the characteristic self-similar revivals remain 
clearly visible, before reaching 
the Page value (dashed line).

The prethermal lifetime $\tau$ scales algebraically as $\tau\sim\theta_z^{-2}$ [Fig.~\ref{Fig3}(b)], consistent with the heating rate expected from Fermi's golden rule~\cite{mori2022heating}. Here $\tau$ is defined as the time at which the EE crosses the threshold $S_{L/2}=8$;
the scaling is insensitive to the chosen threshold (see SM~\cite{supplement}). This long-lived prethermal regime establishes that the predicted self-similar revivals can be resolved experimentally in the presence of generic perturbations.

\textit{Conclusions.}--
In this Letter, we have uncovered nonequilibrium phases of matter exhibiting self-similar quantum revivals using spiral drives.  
These revivals emerge in the time evolution as long as the spiral drive is Gaussian-preserving. 
Their existence relies on an emergent period-6 cycle in the space of transfer matrices, which is unique to the spiral drive and does not arise in previously studied quasiperiodic drives.
Moreover, we have found that the entanglement growth under the spiral drive is tunable between logarithmically slow growth and linear growth, both of which coexist with self-similar revivals. 
When integrability-breaking perturbations are added, despite the eventual heating, the drive can exhibit a long-lived prethermal regime with an algebraically long lifetime, where the characteristic revival survives. Crucially, the spiral drive samples all possible rotation angles in the X-gate, which is thus fundamentally different from any high-frequency limit that has been commonly used to generate prethermalization in driven systems.

The robustness of our findings against generic perturbations also implies that they are experimentally realizable on  quantum simulators, e.g., superconducting qubit~\cite{mi2022time,Liu_2026_quasi,Huang2026ExactCriticalStates}, trapped-ion~\cite{chertkov2022holographic,moses2023race,haghshenas2026digital} and Rydberg-atom platforms~\cite{evered2023high,bluvstein2024logical,manovitz2025quantum}. Our protocol only involves nearest-neighbor Ising coupling gates and single-qubit X-rotations, which are readily implementable with high fidelity on state-of-the-art digital quantum devices. 
We also emphasize that, in addition to measuring entanglement entropy or state fidelity, the self-similar revival can also be diagnosed using the equal-time correlation function; 
see details in End Matter.

Our work draws new connections between nonequilibrium quantum dynamics and the mathematics of quasiperiodic cocycles. A promising future direction is to use this connection to rigorously prove the existence of self-similar revivals for more general classes of drives.
While generic nonintegrable perturbations lead to thermalization and the loss of revivals, an important open question is whether quasiperiodic drives can stabilize self-similar quantum revivals for special classes of nonintegrable perturbations, thereby opening new paths for ergodicity breaking in driven systems.
Finally, going beyond unitary dynamics, an exciting direction is to investigate the interplay between weak measurements and spiral quasiperiodic drives, which could give rise to measurement-induced phase transitions in the memory of the initial state.

\begin{acknowledgments}
\textit{Acknowledgments.}--We thank Marin Bukov, Friedrich Hübner, Roderich Moessner, and Carlo Vanoni for stimulating discussions. B.L. is grateful to Raphaël Krikorian for his help in clarifying the relation of this work to quasiperiodic cocycles.
This work is supported by the Quantum Science and Technology-National Science and Technology Major Project
(No. 2024ZD0301800), by the National Natural Science
Foundation of China (Grant No. 12474214), by ``The Fundamental Research Funds for the Central Universities, Peking University'', and by ``High-performance Computing Platform of Peking University''.
\end{acknowledgments}

\bibliography{ref,QPDriven_ref}

\onecolumngrid 
\section*{End Matter}

\twocolumngrid

\section{Proof of period-6 from reducible cocycles}

\begin{figure}[htp!]
\centering
\includegraphics[width=0.4\textwidth]{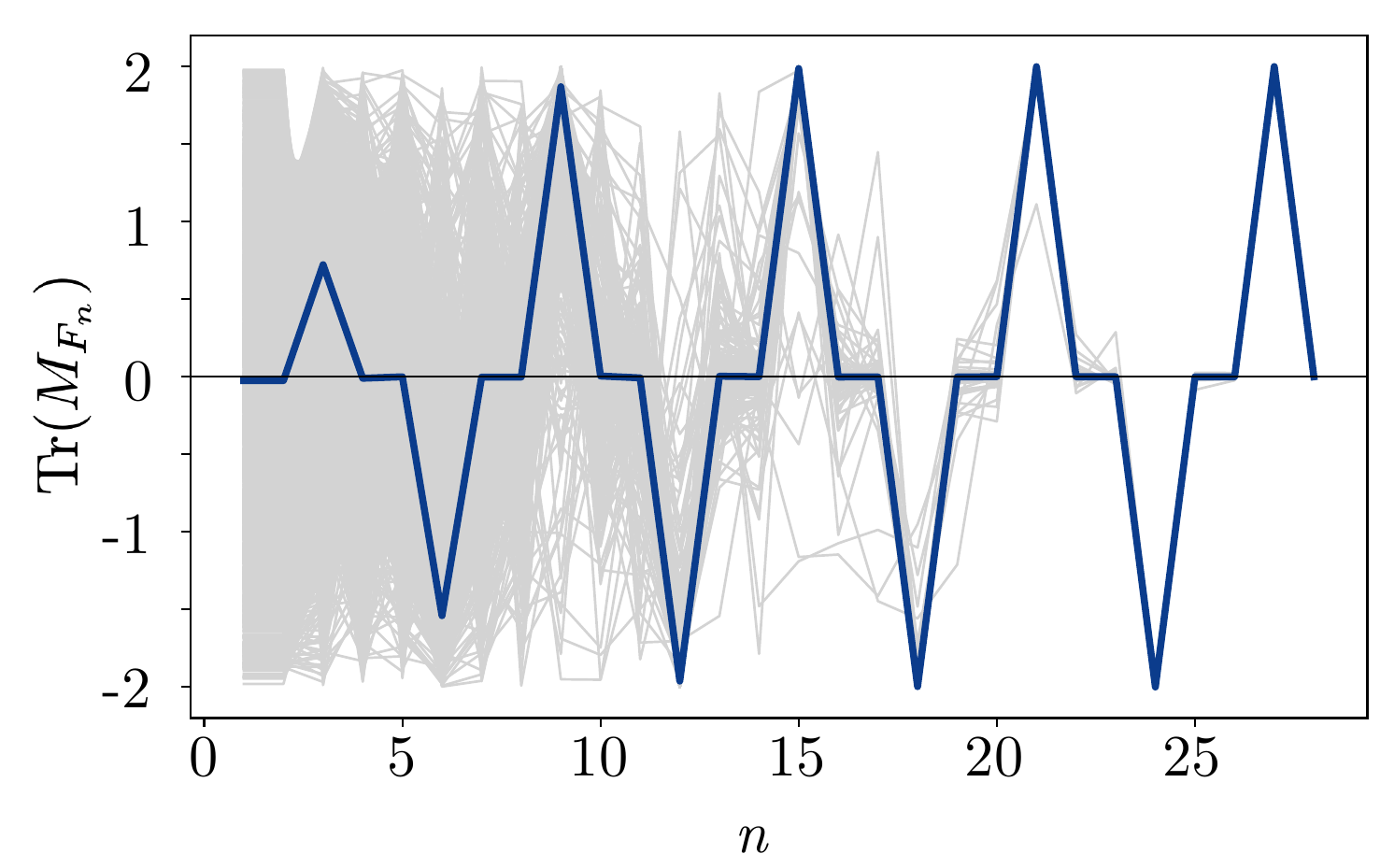}
\caption{\label{figure_app_average} Plot of $\text{Tr}(M_{F_n})$, as defined in~\eqref{eq:matrixproductSU2}, shown as a function of the index $n$ for Haar distributed SU(2) matrices $P$, with 2000 random samples. The averaged value  of the trace is shown in blue, and each realization of $P$ is shown as a gray curve. 
We numerically observe the convergence to the period-6 pattern for all realizations of $P$.}
\end{figure}

The goal of this section is to prove the emergence of a period-6 asymptotic behavior in the matrix product defined by Eq.~\eqref{eq:matrixproductSU2}. Concretely, we will show that the trace $\text{Tr}(M_{F_n})$ converges towards the period-6 orbits, $\{0,0,2,0,0,-2\}$, as a function of $n$, as is numerically observed in Fig.~\ref{figure_app_average}.

In order to prove this analytically, we will make use of the theory of quasiperiodic cocycles.
Specifically, a one-frequency quasiperiodic SU(2) cocycle is a pair \((\varphi,A)\), where
\(\varphi\in\mathbb R\setminus\mathbb Q\) is an irrational rotation, and
\(A:\mathbb T\to SU(2)\) is a smooth map, with \(\mathbb T=\mathbb R/\mathbb Z\).
It defines the discrete evolution
$$
(x,\psi)\mapsto (x+\varphi,A(x)\psi),
\qquad x\in\mathbb T,\quad \psi\in\mathbb C^2.
$$
In the rest of the section, we will consider $A(x)=Pe^{2\pi i x\sigma_z}$, for a nondiagonal matrix $P\in\text{SU}(2)$ and we take $\varphi=(\sqrt{5}-1)/2$.
In this case, the \(n\)-th iteration of the cocycle reads
\[
A_n(x)
=
P e^{2\pi i (x+(n-1)\varphi)\sigma_z}
\cdots
P e^{2\pi i (x+\varphi)\sigma_z}
P e^{2\pi i x\sigma_z}.
\]
Taking \(x=\varphi\), one obtains
\[
A_{F_n}(\varphi)
=
P e^{2\pi i F_n\varphi\sigma_z}
P e^{2\pi i (F_n-1)\varphi\sigma_z}
\cdots
P e^{2\pi i \varphi\sigma_z}.
\]
which coincides with the matrix product defined in~\eqref{eq:matrixproductSU2}, i.e., $M_{F_n}=A_{F_n}(\varphi)$.
A fundamental result in the theory of quasiperiodic cocycles is reducibility, which means that
\begin{equation}
A(x) = B(x+\varphi)CB(x)^{-1},
\end{equation}
for a constant matrix $C\in\text{SU}(2)$ and a quasiperiodic cocycle $B(x)$.
In the rest of this section, we will assume that the quasiperiodic cocycle $A(x)=Pe^{2\pi i x\sigma_z}$ is reducible, which explicitly excludes the case of $P$ being a diagonal matrix. This is a natural assumption, as reducible one-frequency cocycles over SU(2) are globally dense~\cite{Krikorian2001GlobalDO}. A key property of reducible cocycles is that their iteration follows a telescoping identity
\begin{equation}
A_{F_n}(x) = B(x+F_n\varphi)C^{F_n} B(x)^{-1}.
\label{eq:telescopingprod}
\end{equation}
As stated in the main text, from Binet's identity, $F_n\varphi$ converges exponentially fast to $F_{n-1}$. As $B(x)=B(x+1)$, this implies that $B(x+F_n\varphi)$ converges to $B(x)$. 
Then,~\eqref{eq:telescopingprod} converges to a matrix conjugation, which implies that
\begin{equation}
\text{Tr}(A_{F_n}(x)) = \text{Tr}(C^{F_n})+O(\varphi^{n}).
\end{equation}

Our goal is to find the trace properties of the constant matrix $C$, and in particular, show that the trace does not depend on the choice of $P\in\text{SU}(2)$. For this, let us use the symmetry property
\begin{equation}
A(x+\frac{1}{2})=-A(x);
\end{equation}
introducing $K(x)=B\left(x+\frac12\right)^{-1}B(x)$, this condition is equivalent to
\begin{equation}
K(x+\varphi)=-C K(x) C^{-1}.
\label{constrainteq:}
\end{equation}
Up to conjugacy, $C=\text{diag}(e^{i\beta},e^{-i\beta})$, and $K(x)=\begin{pmatrix}
    u(x)&v(x)\\-\bar{v}(x)&\bar{u}(x)
\end{pmatrix}$. Plugging these definitions into~\eqref{constrainteq:} leads to the first constraint
\begin{equation}
u(x+\varphi)=-u(x),
\label{eq:irrationaleqa}
\end{equation}
which implies that $u(x)\equiv0$. Indeed, as $\varphi$ is irrational, iterating~\eqref{eq:irrationaleqa} leads to a dense subset of the circle (recall that $u(x+1)=u(x)$)
where $|u(x)|$ is a constant. In order to get a continuous solution, $u(x)$ has to be constant everywhere. This directly implies that $K$ is off-diagonal, and $|v(x)|=1$. The second constraint is
\begin{equation}
v(x+\varphi)=-e^{2i\beta}v(x).
\end{equation}
This equation can be seen as an eigenvalue equation on the circle, whose solution is $v(x)=ce^{2\pi i k x}$, for $k\in\mathbb{Z}$ and a constant $|c|=1$. This gives rise to an equation for the eigenvalues of $C$,
\begin{equation}
e^{2\pi i k \varphi}=-e^{2i\beta},
\end{equation}
whose solution is $\beta=\pi k \varphi-\frac{\pi}{2}$ for $k\in\mathbb{Z}$. Using the periodicity of $B$, i.e., $B(x+1)=B(x)$, we obtain $K(x+\frac{1}{2})=K(x)^{-1}$. Moreover, we know that $K$ is off-diagonal, such that $K(x)^{-1}=-K(x)$, so $K(x+\frac{1}{2})=-K(x)$. This implies that $k$ is odd, as $e^{\pi i k}=-1$.
We conclude that
\begin{equation}
\text{Tr}(C^{F_n}) = 2\cos(-\frac{\pi F_n}{2}+\pi k F_n\varphi),
\end{equation}
which, using Binet's identity, converges towards
\begin{equation}
\text{Tr}(C^{F_n})\rightarrow 2\cos(-\frac{\pi F_n}{2}+\pi k F_{n-1}),
\end{equation}
and using that $k\in2\mathbb{Z}+1$, it is straightforward to conclude that $\text{Tr}(C^{F_n})$ converges to the values $\{0,0,2,0,0,-2\}$ depending on $n$ mod(6). We thus conclude that at steps $F_{3n}$, $A_{F_{3n}}(x)$ converges to the identity matrix, up to an overall sign.

\section{Dynamics of correlators}

In the main text, we have focused on the time evolution of both fidelity and entanglement entropy for spiral drives. In this section, we provide complementary results on two-point correlators, which can be easily measured on quantum simulators without performing quantum state tomography. Concretely, we will focus on the spin structure factor defined as 
\begin{align}
     S^{xx}(k,n)\equiv\sum_{r}e^{ikr}\langle X_0(n)X_r(n)\rangle,
\end{align}
which can display robust signatures of the self-similar revivals, accessible after a fixed number of cycles in the thermodynamic limit.      
Using the Jordan-Wigner transformation, the structure factor can be obtained using fermionic correlation functions. At momentum $k$, we get 
\begin{align}
    S^{xx}(k,n)=\sum_{r=0}^{L-1} e^{ikr} D_r(n),
\end{align}where
\begin{align}
    D_r(n)= \left\{\begin{matrix}
 1,&r=0 \\
(1-2G_0(n))^2+4(|F_r(n)|^2-|G_r(n)|^2)  &r\neq 0.
\end{matrix}\right.,
\end{align}
with the real-space correlators defined as 
\begin{align}\label{Eq: Grn1}
G_r(n)
&=
\langle c_{j+r}^\dagger c_j \rangle_n
=
\frac{1}{L}
\sum_k
n_k(n) e^{-ikr}\\
F_r(n)
&=
\langle c_{j+r} c_{j} \rangle_n
=
\frac{1}{L}
\sum_k
\kappa_k(n) e^{-ikr},
\end{align}
and the single-particle correlators
\begin{align}\label{Eq:nkkappak1}
n_k(n)&
=
\langle c_k^\dagger c_k \rangle_n
=
\frac{|f_n(k)|^2}{1+|f_n(k)|^2}\\   \kappa_k(n)
&=
\langle c_{-k} c_k \rangle_n
=
\frac{f_n(k)}{1+|f_n(k)|^2}.
\end{align}

\begin{figure}[htp!]
\centering
\includegraphics[width=0.4\textwidth]{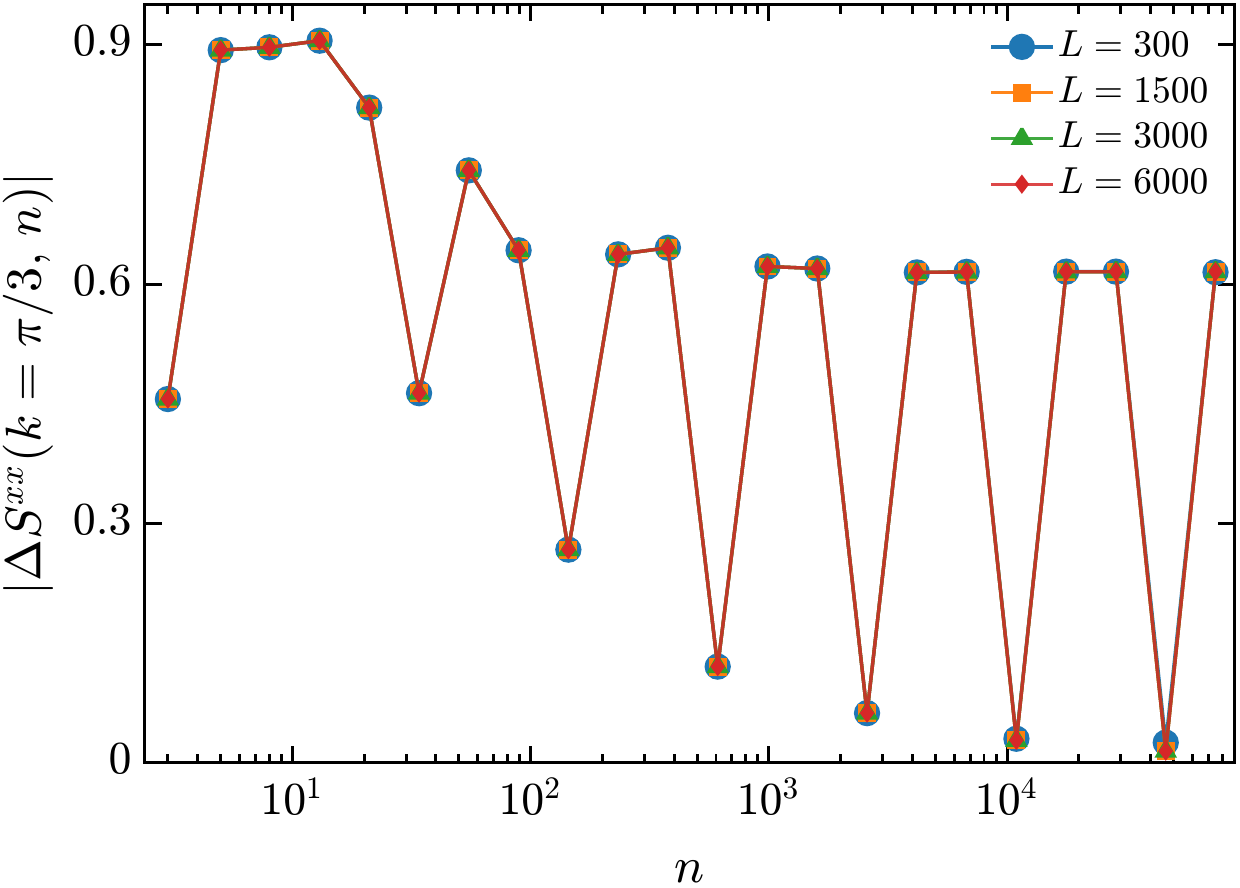}
\caption{Deviation of the structure
factor from its initial value exhibits self-similar revivals. $|\Delta S^{xx}(k=\pi/3,n)| = |S^{xx}(k=\pi/3,n) - S^{xx}(k=\pi/3,0)|$. The locations of the
deep dips are essentially independent of $L$, indicating that the
revivals are intrinsic to the quasiperiodic spiral drive.}
\label{Fig:FSS_Sxx_paper}
\end{figure}

As shown in Fig.~\ref{Fig:FSS_Sxx_paper}, the revivals are highly pronounced for $|\Delta S^{xx}(k=\pi/3,n)|$. We stress that any choice of momentum $k$ would lead to similar revivals. In particular, by increasing the system size $L$, all curves collapse, confirming that it is an intrinsic property of the quasiperiodic drive rather than a finite-size effect.

\clearpage

\renewcommand{\thesection}{S-\arabic{section}} 
\setcounter{section}{0}  
\renewcommand{\theequation}{S\arabic{equation}} 
\setcounter{equation}{0}  
\renewcommand{\thefigure}{S\arabic{figure}} 
\setcounter{figure}{0}  
\renewcommand{\thetable}{S\Roman{table}} 
\setcounter{table}{0}  
\onecolumngrid \flushbottom 

\begin{center}   
	\textbf{\large Supplementary Material for ``Emergent Self-Similar Quantum Revivals in Spiral Drives''}\\
	[1em]
	Xin-Chi Zhou$^{1}$, Liang-Hong Mo$^{2}$, Hongzheng Zhao$^{3}$ and Bastien Lapierre$^{4}$ \\[.1cm]
	{\itshape \small ${}^1$Max Planck Institute for the Physics of Complex Systems, N\"othnitzer Stra\ss e 38, 01187 Dresden, Germany \\ 
    ${}^2$Department of Physics, Princeton University, Princeton, New Jersey, 08544, USA \\ 
    ${}^3$State Key Laboratory of Artificial Microstructure and Mesoscopic Physics, School of Physics, Peking University, Beijing 100871, China\\ 
    ${}^4$Philippe Meyer Institute, Physics Department, \'Ecole Normale Sup\'erieure (ENS), Université PSL, 24 rue Lhomond, F-75231 Paris, France}\\
	(Dated: \today)\\[1cm]
\thispagestyle{titlepage} 
\end{center}

This Supplementary Material contains several appendices providing technical details that support the results presented in the main text:
\begin{enumerate}
    \item In Appendix~\ref{app:gaussianstates}, we review the Gaussian-state formalism used to compute the time evolution of observables.
    \item In Appendix~\ref{app:mechanism}, we prove convergence to the period-6 cycle, both in special cases and through perturbative arguments.
    \item In Appendix~\ref{app:metallic-sequences}, we demonstrate the emergence of revivals for spiral drives defined by different irrational numbers.
    \item In Appendix~\ref{app:floquetapproxx}, we study the quasiperiodic drive using Floquet approximants.
    \item In Appendix~\ref{app:tresholdindep}, we provide further numerical details on the threshold used in the nonintegrable calculations presented in the main text.
\end{enumerate}

\section{Gaussian-state formalism}
\label{app:gaussianstates}
We summarize the Gaussian-state formalism used to compute the
observables in the main text.  Since the spiral drive is quadratic
after the Jordan-Wigner transformation, the many-body dynamics remains
within the fermionic coherent-state manifold and is reduced to
independent momentum sectors.  We first derive the explicit form of M\"obius transformation, and then show how entanglement entropy and fidelity are obtained from the correlation matrix.

\subsection{Unitary evolution of fermionic coherent states}

We first derive the evolution operator for each gate under the coherent-state formalism. The spin Hamiltonian is mapped to a spinless free fermion system after the Jordan-Wigner transformation 
\begin{eqnarray*}
X_j&=&1-2c_j^\dagger c_j,\\
Y_j&=&-i(c_j-c_j^\dagger)\prod_{l<j}(1-2c_l^\dagger c_l),\\
Z_j&=&(c_j+c_j^\dagger)\prod_{l<j}(1-2c_l^\dagger c_l),
\end{eqnarray*}
and we consider the coherent states of the form
\begin{equation}\label{Eq:BCSstate}
    |\psi(\mathcal{N},f)\rangle = \mathcal{N}\prod_{k\in\Lambda_+}\bigl(1+f(k) c_k^\dagger c_{-k}^\dagger\bigr)|0\rangle,
\end{equation}
where $\Lambda_+=\frac{2\pi}{L}\{1/2,\ldots,L/2-1/2\}$ is the set of positive momenta (for antiperiodic boundary conditions), $f(k)$ is a complex amplitude, and $\mathcal{N}$ is a normalization constant. The system remains as a fermionic coherent state after the implementation of Gaussian gates $\mathcal{O}\in\{{\rm ZZ}, {\rm YY}, X\}$~\cite{dreyer2021Quantum,granet2023VolumeLaw}: 
\begin{equation}
    U_{\mathcal{O}}(t) |\psi(\mathcal{N},f(k))\rangle = |\psi(\widetilde{\mathcal{N}},\tilde{f}(k))\rangle,
\end{equation}
with the amplitude updated via a M\"{o}bius transformation 
\begin{equation}
    \tilde{f}(k)=\frac{\alpha_k f(k)+\beta_k}{-\beta_k^* f(k)+\alpha_k^*},
\end{equation}
with $|\alpha_k|^2+|\beta_k|^2=1$. To derive the coefficients $\alpha_k,\beta_k$ systematically, we start from the transverse field Ising Hamiltonian
\begin{equation}
    H = -\sum_j \big(Z_j Z_{j+1} + h X_j\big).
    \label{eq:HIsing}
\end{equation}
Using the Jordan-Wigner transformation, the Hamiltonian becomes
\begin{equation}
    H = -\sum_j\Bigl[c_j^\dagger c_{j+1}+c_{j+1}^\dagger c_j
        -\bigl(c_j c_{j+1}+c_{j+1}^\dagger c_j^\dagger\bigr)\Bigr]
        +2h\sum_j c_j^\dagger c_j.
\end{equation}
After Fourier transformation $c_j=L^{-1/2}\sum_k e^{ikj}c_k$, the Hamiltonian decouples into independent $k>0$ sectors. In the Nambu spinor basis $\Psi_k=(c_k, c_{-k}^\dagger)^T$ one finds $H = \sum_{k>0} 2\Psi_k^\dagger h_k\Psi_k$, with 
\begin{equation}
    h_k=(h-\cos k) \sigma_z+\sin k \sigma_y
\end{equation}
where $\sigma_y,\sigma_z$ are Pauli matrices. Introducing
\begin{equation}
    A_k = h-\cos k,\quad B_k = \sin k,\quad \epsilon_k=\sqrt{A_k^2+B_k^2},
\end{equation}
the time-evolution operator in each $k$ sector is
\begin{equation}
    U_k\equiv e^{ih_k t}
    =\begin{pmatrix}\alpha_k&\beta_k\\-\beta_k^*&\alpha_k^*\end{pmatrix},
    \label{eq:SU2gate}
\end{equation}
with
\begin{equation}
    \alpha_k = \cos(2\epsilon_k t)+i\frac{A_k}{\epsilon_k}\sin(2\epsilon_k t),\qquad
    \beta_k  = \frac{B_k}{\epsilon_k}\sin(2\epsilon_k t).
    \label{eq:alphabeta}
\end{equation}
The gate $U_{\rm ZZ}(\phi_z)=e^{-i\phi_z\sum_j Z_jZ_{j+1}}$ corresponds to~\eqref{eq:HIsing} with $h=0$ and $t=\phi_z$, giving $A_k=-\cos k$, $B_k=\sin k$, $\epsilon_k=1$:
\begin{equation}
    G_{\rm ZZ}(k,\phi_z) =
    \begin{pmatrix}
        \cos(2\phi_z)-i\sin(2\phi_z)\cos k & \sin(2\phi_z)\sin k\\
        -\sin(2\phi_z)\sin k & \cos(2\phi_z)+i\sin(2\phi_z)\cos k
    \end{pmatrix}.
    \label{eq:MZZ}
\end{equation}
Similarly, the gate $U_{\rm YY}(\phi_y)=e^{-i\phi_y\sum_j Y_jY_{j+1}}$ corresponds to~\eqref{eq:HIsing} with $h=0$ and $t=\phi_y$, giving $A_k=-\cos k$, $B_k=-\sin k$, $\epsilon_k=1$ and is given by 
\begin{equation}
    G_{\rm YY}(k,\phi_y) =
    \begin{pmatrix}
        \cos(2\phi_y)-i\sin(2\phi_y)\cos k & -\sin(2\phi_y)\sin k\\
        \sin(2\phi_y)\sin k & \cos(2\phi_y)+i\sin(2\phi_y)\cos k
    \end{pmatrix}.
    \label{eq:MYY}
\end{equation}
The gate $U_X(\theta)=e^{-i\theta\sum_j X_j}$ acts diagonally in momentum space and thus 
\begin{equation}
    G_X(\theta) =
    \begin{pmatrix}e^{2i\theta}&0\\0&e^{-2i\theta}\end{pmatrix}.
    \label{eq:MX}
\end{equation}

\subsection{Time evolution of observables, entanglement entropy and state fidelity}

In this subsection, we explain how the time evolution of the
entanglement entropy and the fidelity are obtained from the evolution matrix $\mathcal{M}_n(k)$ defined in the main text and detailed below. After $n$ driving steps, the evolution in each momentum sector is
represented by
\begin{equation}
\mathcal{M}_n(k)
=
\begin{pmatrix}
\alpha_n(k) & \beta_n(k) \\
-\beta_n^*(k) & \alpha_n^*(k)
\end{pmatrix},
\end{equation}
and the time-evolved amplitude is
\begin{equation}
f_n(k)
=
\frac{
\alpha_n(k) f_0(k) + \beta_n(k)
}{
-\beta_n^*(k) f_0(k) + \alpha_n^*(k)
}.
\end{equation}
For the initial product state used in the main text, one has
$f_0(k)=0$, and therefore
\begin{equation}
f_n(k)
=
\frac{\beta_n(k)}{\alpha_n^*(k)} .
\end{equation}
The Gaussian state
can be characterized by its equal-time
correlation functions.  For each momentum mode, we define
\begin{equation}\label{Eq:nkkappak}
n_k(n)
=
\langle c_k^\dagger c_k \rangle_n
=
\frac{|f_n(k)|^2}{1+|f_n(k)|^2}, \qquad \kappa_k(n)
=
\langle c_{-k} c_k \rangle_n
=
\frac{f_n(k)}{1+|f_n(k)|^2}.
\end{equation}
The corresponding real-space correlators are obtained by Fourier
transforming over the antiperiodic momentum grid,
\begin{equation}\label{Eq: Grn}
G_r(n)
=
\langle c_{j+r}^\dagger c_j \rangle_n
=
\frac{1}{L}
\sum_k
n_k(n) e^{-ikr},
\end{equation}
and 
\begin{equation}\label{Eq:Frn}
F_r(n)
=
\langle  c_{j+r} c_j \rangle_n
=
\frac{1}{L}
\sum_k
\kappa_k(n) e^{-ikr}.
\end{equation}
To compute entanglement entropy, one may construct the Majorana correlation matrix.
We introduce Majorana operators
\begin{equation}
a_{2j-1}
=
c_j+c_j^\dagger,
\qquad
a_{2j}
=
i(c_j-c_j^\dagger).
\end{equation}
For a subsystem $A=[1,\ell]$, the restricted Majorana correlation
matrix $\Gamma_A(n)$ is defined by
\begin{equation}
\langle a_\mu a_\nu\rangle_n
=
\delta_{\mu\nu}
+
i[\Gamma_A(n)]_{\mu\nu},
\qquad
\mu,\nu\in A .
\end{equation}
Translation invariance enforces that $\Gamma_A(n)$ has a block-Toeplitz
form,
\begin{equation}
\Gamma_A(n)
=
[
\Pi_{j-i}(n)
]_{i,j=1,\ldots,\ell},
\end{equation}
with $2\times2$ blocks
\begin{equation}
\Pi_r(n)
=
\begin{pmatrix}
-\varphi_r(n) & \psi_r(n) \\
-\psi_{-r}(n) & \varphi_r(n)
\end{pmatrix},
\end{equation}
where the functions $\varphi_r(n)$ and $\psi_r(n)$ are directly determined
by $f_n(k)$:
\begin{equation}
\varphi_r(n)
=
\frac{i}{2\pi}
\int_{-\pi}^{\pi}
dk 
e^{-ikr}
\frac{
f_n(k)+f_n^*(k)
}{
1+|f_n(k)|^2
},
\end{equation}
and
\begin{equation}
\psi_r(n)
=
\frac{1}{2\pi}
\int_{-\pi}^{\pi}
dk 
e^{-ikr}
\frac{
f_n(k)-f_n^*(k)+|f_n(k)|^2-1
}{
1+|f_n(k)|^2
}.
\end{equation}
The entanglement entropy of the interval $A$ follows from the
eigenvalues $i\nu_p(n)$ of $\Gamma_A(n)$.  Since $\Gamma_A(n)$ is real
antisymmetric, its eigenvalues occur in pairs, with $0\leq \nu_p(n)\leq 1$. The von Neumann entanglement entropy is given by
\begin{equation}
S_A(n)
=
\sum_{p=1}^{\ell}
H_1\!\left(\nu_p(n)\right),
\end{equation}
where 
\begin{equation}
H_1(\nu)
=
-\frac{1+\nu}{2}
\log\frac{1+\nu}{2}
-
\frac{1-\nu}{2}
\log\frac{1-\nu}{2}.
\end{equation}
More generally, the Rényi entropies are given by
\begin{equation}
S_A^{(q)}(n)
=
\sum_{p=1}^{\ell}
H_q\!\left(\nu_p(n)\right),
\end{equation}
with
\begin{equation}
H_q(\nu)
=
\frac{1}{1-q}
\log
\left[
\left(
\frac{1+\nu}{2}
\right)^q
+
\left(
\frac{1-\nu}{2}
\right)^q
\right].
\end{equation}
The fidelity with respect to the initial state can also be calculated.  For two normalized coherent states labeled by
$f(k)$ and $g(k)$ [Eq.~\eqref{Eq:BCSstate}], their overlap factorizes as
\begin{equation}
\langle \psi(f) | \psi(g) \rangle
=
\prod_{k\in\Lambda_+}
\frac{
1+f^*(k)g(k)
}{
\sqrt{
\left(1+|f(k)|^2\right)
\left(1+|g(k)|^2\right)
}
}.
\end{equation}
Thus the fidelity between the initial state and the time-evolved state
is
\begin{equation}
\mathcal{F}(n)
=
|\langle \psi(0)|\psi(n)\rangle|^2
=
\prod_{k\in\Lambda_+}
\frac{
\left|1+f_0^*(k)f_n(k)\right|^2
}{
\left(1+|f_0(k)|^2\right)
\left(1+|f_n(k)|^2\right)
}.
\end{equation}
For the initial product state $f_0(k)=0$, this simplifies to
\begin{equation}
\mathcal{F}(n)
=
\prod_{k\in\Lambda_+}
\frac{1}{1+|f_n(k)|^2}.
\end{equation}
Equivalently, one may compute
\begin{equation}
\log \mathcal{F}(n)
=
-
\sum_{k\in\Lambda_+}
\log\left(1+|f_n(k)|^2\right),
\end{equation}
which is numerically stable when the many-body fidelity becomes
exponentially small in system size.

\section{Self-similar mechanism: proofs in exact and perturbative cases}
\label{app:mechanism}

The main objective of this section is to show that, for any $P=\begin{pmatrix}
    a&b\\-b^*&a^*
\end{pmatrix}\in \mathrm{SU(2)}$ with $|a|^2+|b|^2=1$ and $b\neq0$, the length-$n$ accumulated product
\begin{equation}
M_n(P)  =  D^{n} P D^{n-1} P \cdots D P, \quad
  D \equiv e^{2\pi i\varphi\sigma_z}, \quad \varphi=(\sqrt{5}-1)/2,
\label{eq:matrixproducttoget}
\end{equation}
asymptotically locks into a period-6 cycle along the Fibonacci subsequence
\begin{equation}
    \|M_{F_n}(P)-M_{F_{n+6}}(P)\|\to 0,\quad n\rightarrow \infty.
\end{equation}
Moreover, for $n \equiv 0 \pmod{3}$, $M_{F_n}(P)$ converges to the identity matrix  at the even Fibonacci time steps  up to an overall sign.  The physical consequence of this fact is illustrated in Fig.~\ref{Fig1}(c) in the main text: while all $k$-modes evolve independently, they all converge back to their initial value at steps $F_{3m}$, producing the self-similar revivals visible in both the state fidelity and the entanglement entropy.

We will organize this section as follows:
We first build physical intuition through a spin-echo analogy by analyzing the case in which $P$ is purely off-diagonal in Sec.~\ref{subsec:echo}.
We then show in Sec.~\ref{subsec:perturbationproof} that the revival is stable against small $\mathrm{SU}(2)$ deviations from the off-diagonal case. 
These arguments complement the more general proof shown in the End Matter. In fact, in the rest of the section we do not use any assumption on the reducibility of the cocycle.

Throughout this section we make repeated use of Binet's identity,
\begin{equation}
F_n\varphi=F_{n-1}-(-\varphi)^n,
\label{eq:Binet}
\end{equation}
which controls every decay rate appearing below.

\subsection{Perfect echo limit: Off-diagonal SU(2) element}
\label{subsec:echo}
To build physical intuition through the spin-echo analogy,  we first establish the period-6 orbit for the purely off-diagonal $P(\theta)$ 
\begin{align}
    \label{eq.P_theta}
    P(\theta)=\begin{pmatrix}
    0&e^{i\theta}\\-e^{-i\theta}&0
    \end{pmatrix}.
\end{align}
Notice that $ P(\theta)$ acts as a $\pi$-pulse that flips the $z$-axis of the spin,
\begin{align}\label{Eq:twoindentity}
        P(\theta)\sigma_zP(\theta)^{-1}=-\sigma_z,P(\theta)^2= -\mathbb{I}.
    \end{align}
Consequently, this $\pi$-pulse naturally turns forward precession $D=e^{2\pi i\varphi\sigma_z}$ around the $z$-axis into backward precession 
\begin{align}
        \label{eq.reverse}
        P(\theta)D^j = D^{-j}P(\theta),
    \end{align}
leading to a large simplification in the product structure in Eq.~\eqref{eq:matrixproducttoget}
\begin{align}\label{Eq:AAD}
       D^j P(\theta)D^{j-1}P(\theta)=-D.
\end{align}
Thus, each neighboring pair of pulses produces a perfect echo up to a residual rotation $-D$.
Accumulating all the residual rotations  over the full product in  Eq.~\eqref{eq:matrixproducttoget} yields the closed form
\begin{align}
        M_n\big( P(\theta)\big)&= \left\{\begin{matrix}
(-D)^{n/2}  &n \,  \mathrm{even} \\
(-1)^{(n-1)/2}D^{(n+1)/2}P(\theta)  &n  \, \mathrm{odd} .
\end{matrix}\right.\\
&= \exp(2\pi i \varphi \left\lceil n/2 \right\rceil\sigma_z)P(\theta)^{n},\label{eq:closedformula}
    \end{align}
where we have used the identity
    \begin{align}
        P(\theta)^{n} =  (-1)^{\lfloor n/2\rfloor} P(\theta)^{n \mathrm{mod}  2},
    \end{align}
which follows directly from Eq.~\eqref{Eq:twoindentity}. At the Fibonacci length, Eq.~\eqref{eq:closedformula} cleanly separates the $n$-dependence of $ M_{F_n}\big( P(\theta)\big)$ into a discrete factor $P(\theta)^{F_n}$ and a continuous factor $\exp(2\pi i\varphi\lceil F_n/2\rceil\sigma_z)$\footnote{$\lceil x\rceil$ and $\lfloor x\rfloor$ denote the ceiling and
floor functions, i.e.\ the smallest integer $\ge x$ and the largest integer
$\le x$, respectively.}.

The discrete factor is governed by $F_n\bmod 4$. The Fibonacci sequence modulo $4$ has period $6$, with $F_n\bmod 4=\{0,1,1,2,3,1\}$ for $n\bmod 6\in\{0,1,\ldots,5\}$, so that
\begin{equation}
    P(\theta)^{F_n}=\{\mathbb{I},     P(\theta),     P(\theta), -\mathbb{I}, -    P(\theta),     P(\theta)\}\qquad(n\bmod 6).
\end{equation}

The continuous factor is controlled by Binet's identity~\eqref{eq:Binet}. Substituting it into the exponent gives
\begin{equation}
    \exp\bigl(2\pi i\varphi\lceil F_n/2\rceil\sigma_z\bigr)\xrightarrow{n\to\infty}\{-\mathbb{I},e^{i\pi\varphi\sigma_z},e^{i\pi\varphi^{-1}\sigma_z}\}\qquad(n\bmod 3),
\end{equation}
with a residual error of order $\varphi^n$ that decays exponentially since $\varphi=(\sqrt{5}-1)/2<1$.

Combining the two factors, $M_{F_n}(P(\theta))$ converges to the period-6 cycle
\begin{equation}
\label{eq:revivalclaim}
M_{F_n}\bigl(P(\theta)\bigr) \longrightarrow \{-\mathbb{I},  P(\theta+\pi\varphi),  P(\theta+\pi\varphi^{-1}), \mathbb{I}, -P(\theta+\pi\varphi),  P(\theta+\pi\varphi^{-1})\}.
\end{equation}
In particular, at the Fibonacci steps $n=3m$ the product approaches $\pm\mathbb{I}$, which directly leads to the revival.

\subsection{Perturbation proof for deviations from the pure off-diagonal case}
\label{subsec:perturbationproof}
We further show that the revival $M_{F_n}\to\pm\mathbb{I}$ at $n\equiv 0\pmod{3}$ is stable when the off-diagonal $P(\theta)$ of Eq.~\eqref{eq.P_theta} is replaced by an $\mathrm{SU}(2)$ matrix with a small diagonal element, 
\begin{equation}
    P=\begin{pmatrix}\epsilon & \beta\\ -\beta^* & \epsilon^*\end{pmatrix},\qquad |\epsilon|^2+|\beta|^2=1,\quad |\epsilon|\ll 1.
    \label{eq:Pperturbed}
    \end{equation}
Concretely, we will show that the resulting first-order correction to $M_n(P)$ decays as $\varphi^{n}$ along the even-Fibonacci subsequence.

In the perturbation~\eqref{eq:Pperturbed} we have $|\beta|=\sqrt{1-|\epsilon|^2}=1+O(|\epsilon|^2)$. Writing $\beta=|\beta|e^{i\theta}$ and introducing $R_z(\theta/2)\equiv e^{i(\theta/2)\sigma_z}$, the phase of the off-diagonal entry can be absorbed into a global $z$-rotation,
\begin{equation}
P'\equiv R_z(-\theta/2) P R_z(\theta/2) = P_0 + \epsilon_1 \mathbb{I} + i\epsilon_2 \sigma_z+O(|\epsilon|^2),
\label{eq:reducedP}
\end{equation}
where $\epsilon_1=\mathrm{Re} \epsilon$, $\epsilon_2=\mathrm{Im} \epsilon$, and $P_0\equiv i\sigma_y=P(0)$ is in the off-diagonal family~\eqref{eq.P_theta}. Since $[R_z,D]=0$, we have
\begin{equation}
M_n(P)=R_z(\theta/2)M_n(P')R_z(-\theta/2).
\label{eq:conjugationreduction}
\end{equation}
The two operators therefore share the same dynamics up to a $R_z$ rotation, and it is enough to analyse the canonical perturbation~\eqref{eq:reducedP}, i.e.\ deviations of $P_0$ along $\mathbb{I}$ and along $i\sigma_z$.
Importantly, to the leading order perturbation of~\eqref{eq:reducedP}, $\epsilon_1,\epsilon_2$ contribute to the deviation linearly and independently in these two directions. Therefore, 
in the following we will consider their accumulated perturbative effects separately and show both of them vanish along the even-Fibonacci subsequence.

\subsubsection{First-order expansion for the $\mathbb{I}$-direction}

We consider first the $\mathbb{I}$-direction.  Defining $A_j\equiv D^{j}P_0$ , each layer in Eq.~\eqref{eq:matrixproducttoget} becomes $D^{j}(P_0+\epsilon_1 \mathbb{I})= A_j + \epsilon_1D^{j}$. Inserting a single error at the $s$-th layer and leaving every other layer ideal contributes
\begin{equation}
A_{F_n}\cdots A_{s+2} D^{s+1} A_{s}\cdots A_1
=M_{F_n}(P_0) \bigl[M_{s}(P_0)\bigr]^{-1} P_0^{-1} M_{s}(P_0).
\end{equation}
 Summing over the position $s$ of the inserted error yields, to first order in $\epsilon_1$,
\begin{equation}
M_{F_n}(P_0+\epsilon_1\mathbb{I})
 = M_{F_n}(P_0) \Bigl[ \mathbb{I} + \epsilon_1\sum_{s=0}^{F_n-1}K_s \Bigr] + O(\epsilon_1^2),
\label{eq:perturbationseries}
\end{equation}
with
\begin{align}
    K_{s}\equiv [M_s(P_0)]^{-1}P_0^{-1}M_{s}(P_0).
\end{align}

Substituting $M_s(P_0)$ from~\eqref{eq:closedformula} (at $\theta=0$) and using $P_0^{-1}=-P_0$ together with~\eqref{Eq:twoindentity}, one has
\begin{equation}
K_{2q}=-D^{-2q}P_0,\qquad K_{2q+1}=-D^{2q+2}P_0.
\label{eq:Ks_explicit}
\end{equation}
At $F_n=2q$, we have
\begin{align}
\sum_{s=0}^{F_n-1}K_s&=\sum_{k=0}^{q-1}K_{2k}+\sum_{k=0}^{q-1}K_{2k+1}\nonumber\\
&=-\sum_{k=0}^{q-1}D^{-2k}P_0  -D^{2}\sum_{k=0}^{q-1}D^{2k}P_0\nonumber\\
&= -D^{-2(q-1)}\frac{\mathbb{I}-D^{2F_n}}{\mathbb{I}-D^{2}}P_0.
\label{eq:Ksum_even}
\end{align}
Since $\varphi$ is irrational, $(\mathbb{I}-D^{2})^{-1}$ is bounded and the prefactor $D^{-2(q-1)}$ has unit norm, so the size of the sum is set entirely by the numerator $\mathbb{I}-D^{2F_n}$.  Substituting Binet's identity~\eqref{eq:Binet} into the exponent gives
\begin{equation}
D^{2F_n} = e^{4\pi i\varphi F_n\sigma_z} = e^{4\pi i F_{n-1}\sigma_z} e^{-4\pi i(-\varphi)^n\sigma_z} = e^{-4\pi i(-\varphi)^n\sigma_z} \longrightarrow \mathbb{I},
\end{equation}
so that $\|\mathbb{I}-D^{2F_n}\|=O(\varphi^n)$. Combined with~\eqref{eq:perturbationseries}, this implies, on the even-Fibonacci subsequence ($n\equiv 0\pmod 3$),
\begin{equation}
M_{F_n}\bigl(P_0+\epsilon_1\mathbb{I}\bigr)
 = M_{F_n}(P_0) \bigl[ \mathbb{I}+O(\epsilon_1\varphi^n) \bigr] \longrightarrow \pm\mathbb{I}.
\end{equation}

\subsubsection{First-order expansion for the $\sigma_z$-direction}
The perturbation $P_0\mapsto P_0+i\epsilon_2\sigma_z$ at each layer gives $D^{j}P=A_j+i\epsilon_2\sigma_zD^{j}$, since $[D,\sigma_z]=0$. Expanding to first order and using the anticommutation $\sigma_z A_j=-A_j\sigma_z$, 
\begin{align}
    M_{F_n}(P_0+i\epsilon_2\sigma_z)
 = M_{F_n}(P_0) \Bigl[ \mathbb{I}+i\epsilon_2 
\Bigl(\sum_{s=0}^{F_n-1}(-1)^{s}K_s\Bigr)\sigma_z \Bigr] + O(\epsilon_2^2).
\label{eq:MepsZ}
\end{align}
Evaluating the alternating sum at even $F_n$ gives
\begin{equation}
\sum_{s=0}^{F_n-1}(-1)^{s}K_s
 = \frac{D^{2-F_n}(D^{F_n}-\mathbb{I})^{2}}{D^{2}-\mathbb{I}} P_0,
\label{eq:altsum_even}
\end{equation}
where we used the factorization $D^{F_n+2}-2D^{2}+D^{2-F_n}=D^{2-F_n}(D^{F_n}-\mathbb{I})^{2}$. The numerator is now quadratic in $\mathbb{I}-D^{F_n}$, so the same Binet estimate $\|\mathbb{I}-D^{F_n}\|=O(\varphi^n)$ yields
\begin{equation}
\Bigl\|\sum_{s=0}^{F_n-1}(-1)^{s}K_s\Bigr\| = O(\varphi^{2n}).
\end{equation}

Putting the two directions together, the accumulated operator generated by the general perturbation~\eqref{eq:Pperturbed} satisfies
\begin{equation}
M_{F_n}(P) = \pm\mathbb{I} + O(\varphi^n) + O\bigl(\epsilon_1 \varphi^{n}\bigr) + O\bigl(\epsilon_2 \varphi^{2n}\bigr)
\end{equation}
on the even-Fibonacci subsequence ($n\equiv 0\pmod 3$). This establishes the revival at first order in the perturbation:  $M_{F_n}\to\pm\mathbb{I}$ survives at $n\equiv 0\pmod 3$, with exponential convergence rate (at least) $\varphi^{n}$ in $n$, equivalently $\varphi^{6}$ per six Fibonacci steps.

\section{Spiral drives from various irrational numbers}
\label{app:metallic-sequences}

In the main text, we have focused our attention on the spiral drive generated by the golden ratio, \(\varphi=(\sqrt{5}-1)/2\), and found revival dynamics at special times associated with Fibonacci numbers  \(F_m\).  We now show that the self-similar quantum revivals are not unique to the golden-ratio case. Rather, they appear for a broad class of irrational
frequencies whose rational approximants are generated by recursive
integer sequences.  

\subsection{Spiral drive from metallic ratios}
A direct generalization of the golden ratio is provided by the metallic ratios. 
We denote
the metallic family by a positive integer \(\kappa\), and the recurrence relation of the metallic sequence is given by 
\begin{equation}
    F^{(\kappa)}_m = \kappa F^{(\kappa)}_{m-1} + F^{(\kappa)}_{m-2}, \qquad m\geq 2 .
    \label{eq:metallic_sequence}
\end{equation}
with $F^{(\kappa)}_0 = 0$ and $F^{(\kappa)}_1 = 1$.
The corresponding metallic  irrational frequency  is
\begin{align}
    \alpha_\kappa
    =
    \lim_{m\to\infty}
    \frac{F^{(\kappa)}_{m-1}}{F^{(\kappa)}_m}
    =
    \frac{2}{\kappa+\sqrt{\kappa^2+4}} .
    \label{eq:metallic_alpha}
\end{align}
For \(\kappa=1\), Eqs.~\eqref{eq:metallic_sequence}
and~\eqref{eq:metallic_alpha} reduce to the usual Fibonacci sequence
and the golden ratio used in the main text.  For
\(\kappa=2,3,4\), they give the silver, bronze, and fourth metallic
irrational frequencies, respectively.

\begin{figure*}[htp!]
\centering
\includegraphics{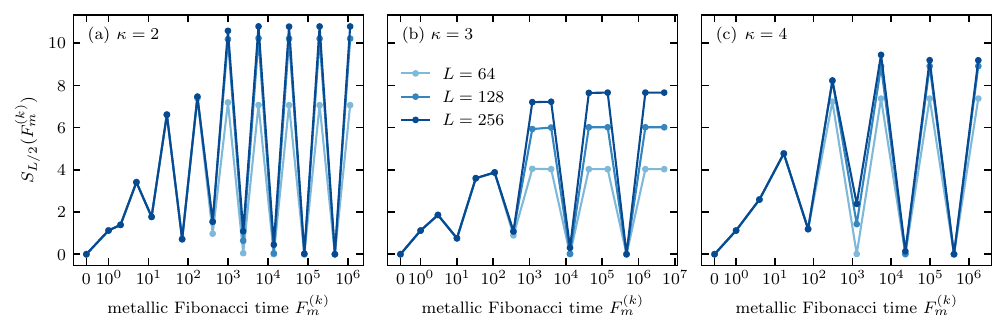}
\caption{\label{Fig:MetallicNumber} 
Half-chain entanglement entropy $S_{L/2}(n)$ for spiral drives generated by metallic
irrational frequencies. $S_{L/2}(n)$ is measured at the metallic times $n=F_m^{\kappa}$, for
\(\kappa=2,3,4\), respectively. The self-similar revivals appear beyond the golden-ratio case. After the revival time, the entanglement entropy further exhibits a parity-dependent pattern: odd
\(\kappa\) gives a period-3 sequence with two plateau-like values and
one revival, while even \(\kappa\) gives a period-two alternation between
plateaus and revivals. Here we fix $\phi_z=\pi/3$.}
\end{figure*}

The revival times are organized by the metallic sequence
\(F^{(\kappa)}_m\). As shown in Fig.~\ref{Fig:MetallicNumber}, the
half-system entanglement entropy exhibits pronounced revival dips
when the time evolution is measured at these metallic times $n=F^{(\kappa)}_m$. In addition to the revival dips, the sequence of
entanglement values displays a simple parity-dependent structure. For
odd \(\kappa\), the entanglement entropy exhibits a period-3 pattern, consisting of
two plateau-like values followed by one revival. For even \(\kappa\), the
pattern reduces to a period-two alternation between plateau-like values
and revivals. This distinction originates from the intrinsic parity
structure of the metallic sequences. In particular, the argument used in
Sec.~\ref{app:mechanism} to explain the period-6 revival structure of
the Fibonacci case can be generalized directly to the metallic sequences
with arbitrary \(\kappa\).

\begin{figure}[htp!]
\centering
\includegraphics[width=0.7\textwidth]{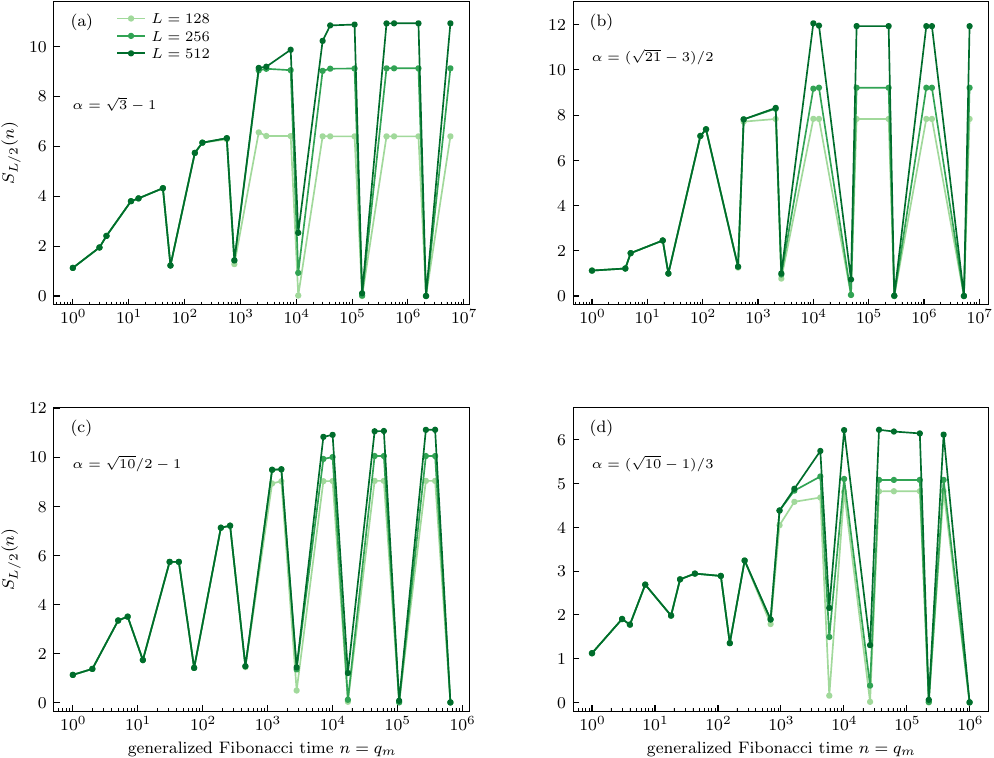}
\caption{\label{Fig:continuedFraction} 
Time evolution of the half-system entanglement entropy for various spiral drives generated by non-metallic periodic continued fractions. The entanglement entropy is measured at the convergent times \(n=q_m\), where \(q_m\) is the
denominator of the \(m\)th continued-fraction convergent. The four panels correspond to
representative quadratic irrational frequencies:
\([0;\overline{1,2}]\) in (a), \([0;\overline{1,3}]\) in (b),
\([0;\overline{1,1,2}]\) in (c), and
\([0;\overline{1,2,1}]\) in (d). In all cases, the entanglement displays both plateau-like values and revival dips at the convergent times. Here we fix
\(\phi_z=\pi/3\).
}
\end{figure}

\subsection{Spiral drive from quadratic irrational numbers}

The revival dynamics can similarly be observed for a more general class of irrational numbers called quadratic irrational numbers. These irrational numbers are characterized by a periodic continued fraction expansion and contain the previously studied metallic ratios as special cases. 
Specifically, given an irrational number $\alpha$, its continued-fraction expansion is defined as
\begin{equation}
    \alpha=[a_0;a_1,a_2,a_3,\ldots]\equiv a_0 + \frac{1}{a_1 + \frac{1}{a_2 + \frac{1}{a_3 + \cdots}}},
\end{equation}
where $a_0\in\mathbb{Z}$ and $a_n\in\mathbb{Z}_{>0}$ for $n\geq 1$.
The $n$th convergent is defined as the rational number obtained by cutting off the infinite continued fraction after $a_n$, i.e.,
\begin{equation}
    \frac{p_n}{q_n}=[a_0;a_1,\ldots,a_n].
\end{equation}
In particular, the numerator and the denominator are generated by
\begin{equation}
    p_n = a_n p_{n-1}+p_{n-2}, \qquad
    q_n = a_n q_{n-1}+q_{n-2},
\end{equation}
with initial conditions $p_{-2}=0$, $p_{-1}=1$, $q_{-2}=1$, and $q_{-1}=0$. The integers \(q_n\) thus provide the natural generalization of the
Fibonacci steps, and we refer to them as the
convergent times. Periodic continued fractions are then obtained by repeating the finite number of coefficients indefinitely, i.e.,
\begin{equation}
    \alpha=[0;\overline{a_1,a_2,\ldots,a_p}]
    =[0;a_1,a_2,\ldots,a_p,a_1,a_2,\ldots],
\label{eq:irrationalnumerperiodic}
\end{equation}
or equivalently,
\begin{equation}
    \alpha=
    \cfrac{1}{a_1+\cfrac{1}{a_2+\cfrac{1}{\cdots+
    \cfrac{1}{a_p+\cfrac{1}{a_1+\cdots}}}}}.
\end{equation}
Thus, quadratic irrational numbers are determined by the repeated coefficient sequence itself. Metallic ratios form the simplest, period-one subclass of such periodic  
continued fractions, as they read
\([0;\overline{\kappa}]\). In this case, the convergent times reduce precisely to the metallic sequence \(F_m^{(\kappa)}\).

\begin{figure*}[htp!]
\centering
\includegraphics{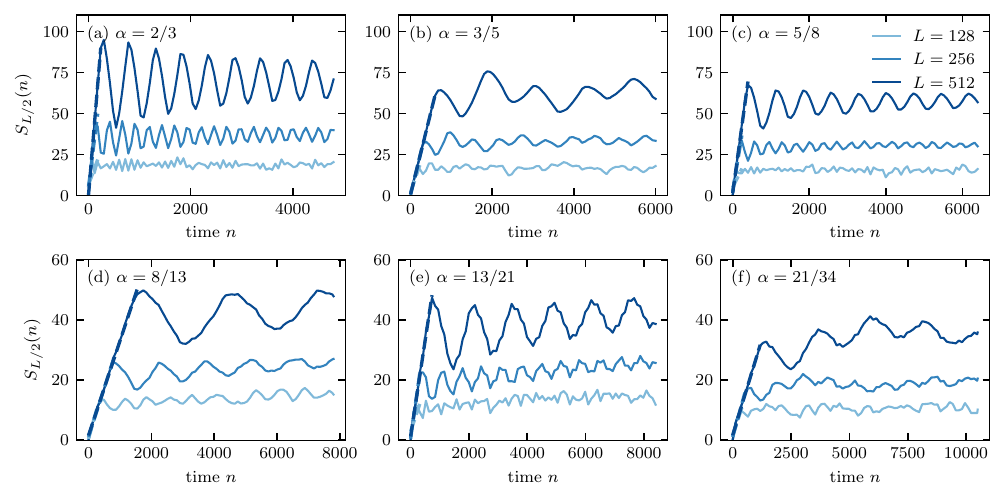}
\caption{\label{Fig:rational_EE} 
Floquet approximants to the spiral drive.
The rational drives are obtained by replacing the irrational rotation
frequency by $\alpha=F_m/F_{m+1}$.
The half-system entanglement entropy $S_{L/2}(n)$ initially grows linearly for each rational approximation; dashed lines
indicate the linear fits used to extract the linear-growth coefficient
$v_m$.}
\end{figure*}

\begin{figure}[htp!]
\centering
\includegraphics{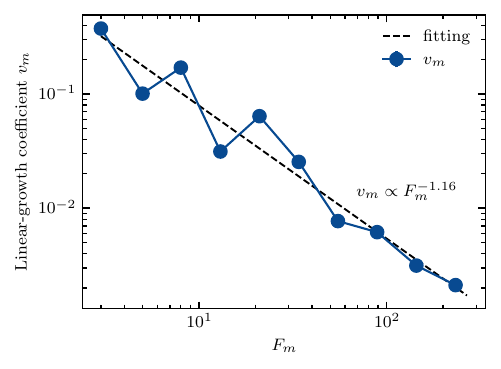}
\caption{\label{Fig:velocityScaling} 
Suppression of the linear-growth coefficient in the irrational limit.
The blue markers show the coefficient \(v_m\), extracted from the
initial linear regime of the half-system entanglement entropy
\(S_{L/2}(n)\), as a function of the Fibonacci scale \(F_m\) of the
rational approximant.
The dashed line denotes a power-law fit,
\(v_m \propto F_m^{-1.16}\). The decay of \(v_m\) shows that the transient linear-growth
regime of the periodic approximants vanishes as the quasiperiodic
spiral-drive limit is approached, consistent with the results shown in the main text.}
\end{figure}

Fig.~\ref{Fig:continuedFraction} shows the half-chain entanglement entropy at convergent
times \(n=q_m\) for various choices of periodic irrational numbers. These times play the same role as the Fibonacci or
metallic times in the previous subsection. In fact, when the time evolution is
sampled at \(n=q_m\), the half-chain entanglement entropy exhibits
recurrent plateau-like values and sharp revival dips.
These results demonstrate that the self-similar revival structure is a generic feature of spiral drives generated by quadratic irrational frequencies and thus extends far beyond the case of the golden ratio.

\section{Floquet approximants and the suppression of linear entanglement growth}
\label{app:floquetapproxx}
In this section, we use Fibonacci rational approximants to systematically approach the quasiperiodic spiral drive from a sequence of periodic Floquet drives, thereby tracking how the ballistic entanglement growth characteristic of Floquet dynamics is progressively suppressed in the quasiperiodic limit. To approach the irrational rotation in a controlled way, we replace the golden-ratio angle by its Fibonacci rational approximants $\alpha_m=\frac{F_m}{F_{m+1}}$, which converge to the irrational value in the limit $m\to\infty$. For any finite $m$, however, $\alpha_m$ is rational, so the sequence of $X$-gate rotations is periodic and the corresponding dynamics defines a Floquet drive with a finite period. This construction allows us to systematically track how the entanglement dynamics evolves from periodic Floquet approximants to the true quasiperiodic spiral drive. As shown below, the coefficient of the initial linear entanglement growth $v_m$ decreases with increasing $m$ and vanishes upon approaching the irrational limit, consistent with the anomalously slow entanglement dynamics of the spiral drive discussed in the main text.

In Fig.~\ref{Fig:rational_EE}, we show the time evolution of the half-chain entanglement entropy for successive Fibonacci rational approximants $\frac{F_m}{F_{m+1}}$. For each finite $m$, the drive is periodic and the entanglement entropy exhibits an initial linear growth, as expected from ballistic propagation of quasiparticles. However, the slope of this linear growth progressively decreases as the rational approximant approaches the irrational spiral drive.

Figure~\ref{Fig:velocityScaling} quantifies the suppression of the linear-growth coefficient \(v_m\), extracted from the transient linear
growth of the entanglement entropy for each rational approximation. A power-law fit gives
\(v_m\propto F_m^{-1.16}\), indicating that the linear-growth coefficient systematically decreases as the sequence of periodic drives
approaches the quasiperiodic spiral drive. Since \(F_m\to\infty\) in
the quasiperiodic limit, this scaling implies that the transient
linear-growth coefficient vanishes in the spiral drive.  This rational-approximation analysis therefore provides a complementary numerical diagnosis of
the anomalously slow entanglement growth observed in the main text.

\section{Threshold-independent scaling of the prethermal lifetime}
\label{app:tresholdindep}

\begin{figure}[htp!]
\centering
\includegraphics[width=\textwidth]{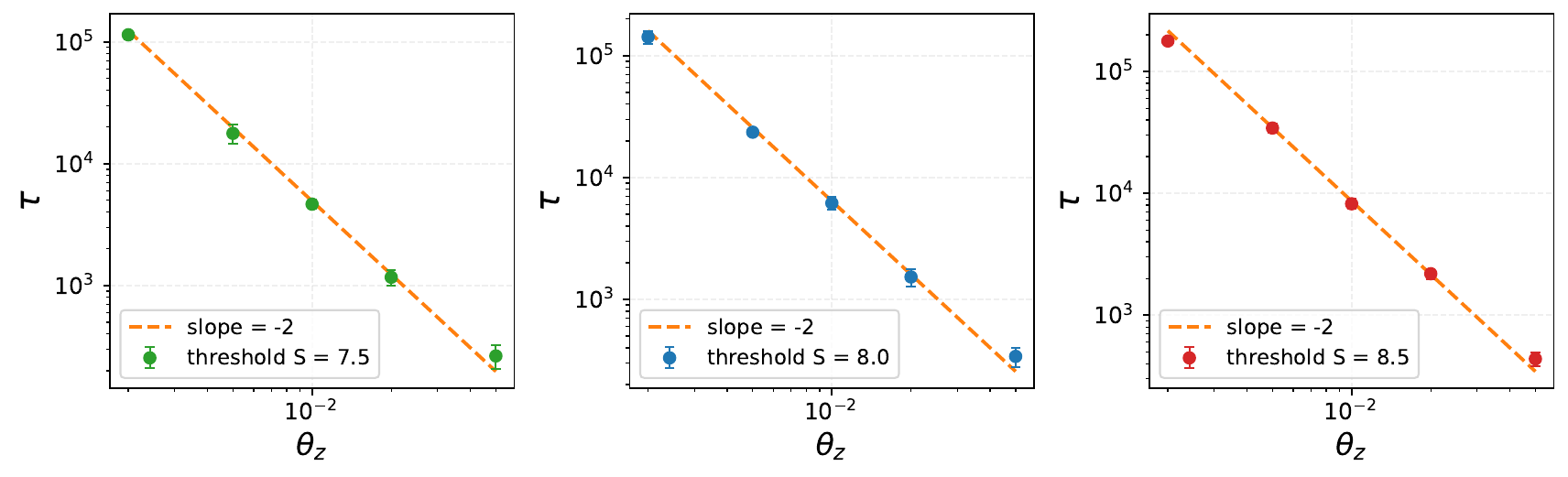}
\caption{Lifetime $\tau$ versus drive amplitude $\theta_z$ extracted at three entanglement entropy thresholds: (a) $S=7.5$,  (b) $S=8.0$, (c) $S=8.5$. Dashed lines show the $\tau\propto\theta_z^{-2}$ scaling. Error bars are $1\sigma$ across realizations.}

\label{Fig:EE_lifetime_threshold_scan}
\end{figure}

In the main text, the prethermal lifetime \(\tau\) is extracted from the time at which the half-chain entanglement entropy crosses a fixed threshold, chosen as \(S_{L/2}=8\). Here we verify that the resulting scaling does not depend on the particular threshold choice. For each noise amplitude \(\theta_z\), we compute the entanglement dynamics for different disorder realizations and define the lifetime $\tau$ as the first time at which the entanglement is above the threshold.  We repeat the analysis for three different thresholds $S=\{7.5, 8.0, 8.2\}$. As shown in Fig.~\ref{Fig:EE_lifetime_threshold_scan}, all three choices give the same algebraic scaling,
\begin{equation}
    \tau\propto \theta_z^{-2}.
\end{equation}
This demonstrates that the exponent is robust and does not depend on the specific operational definition of the lifetime.

\end{document}